\RequirePackage{fix-cm}
\documentclass[twocolumn]{svjour3}       
\smartqed  
\usepackage{graphicx}
\usepackage{url}
\begin{document}

\title{A survey of comics research in computer science}


\author{Olivier Augereau \and Motoi Iwata \and Koichi Kise}


\institute{F. Author \at
              first address \\
              Tel.: +123-45-678910\\
              Fax: +123-45-678910\\
              \email{fauthor@example.com}           
           \and
           S. Author \at
              second address
}

\date{Received: date / Accepted: date}

\maketitle

\begin{abstract}
Graphical novels such as comics and mangas are well known all over the world.
The digital transition started to change the way people are reading comics, more and more on smartphones and tablets and less and less on paper.
In the recent years, a wide variety of research about comics has been proposed and might change  the way comics are created, distributed and read in future years.
Early work focuses on low level document image analysis: indeed comic books are complex, they contains text, drawings, balloon, panels, onomatopoeia, etc.
Different fields of computer science covered research about user interaction and content generation such as multimedia, artificial intelligence, human-computer interaction, etc. with different sets of values.
We propose in this paper to review the previous research about comics in computer science, to state what have been done and to give some insights about the main outlooks.
\end{abstract}

\section{Introduction}

Research on comics have been done independently in several research fields such as document image analysis, multimedia, human-computer interaction, etc. with different sets of values.
We propose to review the research of all of these fields and to organize them in order to understand what is possible to do about comics with the state of the art methods.
We also give some ideas about the future possibility of comics research.

We introduced a brief overview of comics research in computer science~\cite{augereau2017overview} during the second edition of the international workshop on coMics ANalysis, Processing and Understanding (MANPU).
The first edition of MANPU workshop took place during ICPR 2016 (International Conference on Pattern Recognition) and the second one took place during ICDAR 2017 (International Conference on Document Analysis and Recognition).
It shows that comics can interest a large variety of researchers from pattern recognition to document analysis.
We think that the multimedia and interface communities could have some interest too, so we propose to present the research about comics analysis with a broader view.

In the next part of the introduction we will explain the importance of comics and its impact on the society with a brief overview of the open problems.

\subsection{Comics and society}
Comics in the USA, mangas in Japan or bandes dessin\'ees in France and Belgium are graphic novels which have a worldwide audience.
They are respectively an important part of the American, Japanese and Francophone cultures. They are often considered as a soft power of these countries, especially mangas for Japan~\cite{lam2007japan,hall2013struggle}.
In France, bandes dessin\'ee is considered as an art, and is commonly refereed as the ``ninth art''~\cite{Screech2005} (as compared to cinema which is the seventh art).

However, several years ago it was not the case.
Comics was considered as ``children literature'' or ``sub-literature'' as it contains a mixture of images and text.
But more lately comics got a great deal of interest when people recognized it as a complex form of graphic expression that can convey deep ideas and profound aesthetics \cite{Christiansen2000}.

The market of comics is large.
According to a report published in February 2017 by ``The All Japan Magazine and Book Publisher's and Editor's Association'' (AJPEA), the sale of mangas in Japan represents 445.4 billion yens (around 4 billion dollars) in 2016\footnote{\url{http://www.ajpea.or.jp/information/20170224/index.html}}.
In this report, we can see that the market is stable between 2015 and 2014, but a large progression of the digital market can be observed: it almost doubled from 2014 to 2016.
The digital format has several advantages for the readers: it can be displayed on smartphones or tablets and be read anytime, anywhere.
For the editors, the cost of publication and distribution is much lower as compared to the printed version.

However, even if the format changed from paper to screen, no added value has been proposed to the customer.
We think that the democratization of digital format is a good opportunity for the researchers from all computer science fields to propose new services such as augmented comics, recommendation systems, etc.   

\begin{figure}[t]
  \centering
  \includegraphics[width=0.95\linewidth]{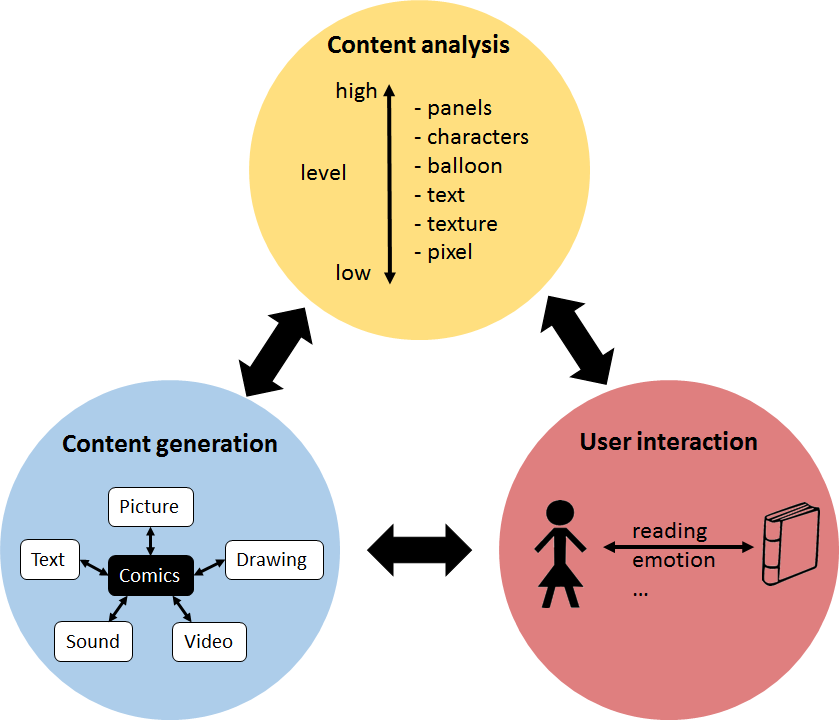}
  \caption{We arranged the comics research into three inter-dependent categories: 1) content analysis, 1) content generation, and 3) user interaction. }
  \label{map}
\end{figure}

\subsection{Research and open problems}
The research about comics is quite challenging because of the nature of this medium.
Comics contains a mixture of drawings and text.
To fully analyze and understand the content of comics, we need to consider natural language processing to understand the story and the dialogues; and computer vision to understand the line drawings, characters, locations, actions, etc.
A high-level analysis is also necessary to understand events, emotions, storytelling, the relations between the characters, etc.
A lot of related research has been done for covering similar aspect for the case of natural images (i.e. photographic imagery) and videos by classic computer vision.
However, the hight variety of drawings and low availability of labeled dataset make the task harder than natural images.

We organized the research about comics in the three following main categories, as illustrated in Fig.~\ref{map}:
\begin{enumerate}
\item content analysis: getting information about raw images and extracting from high to low-level structured descriptions,
\item content generation: comics can be used as an input or output to generate new contents. Content conversion and augmentation are possible from comics to comics, comics to other media, other media to comics;
\item user interaction: analyzing human reading behavior and internal states (emotions, interests) based on comics contents, and reciprocally, analyzing comics contents based on human behavior and interactions.
\end{enumerate}

Research about comics in computer science has been done covering several aspects but is still an emerging field.
Much research has been done by researchers from the DIA (Document Image Analysis) and AI (Artificial Intelligence) communities and focuses on content analysis, understanding, and segmentation.
Another part of the research is addressed by graphics and multimedia communities and consists in generating new contents or enriching existing contents such as adding colors to black and white pages, creating animation, etc.
The last aspect concerns the interaction between users and comics which is mainly addressed by the HCI (Human-Computer Interaction) researchers.
All these three parts are inter-dependent: segmenting an area of a comic page is important if we want to manipulate and modify it, or if we want to know which area the user is interacting with.
Analyzing the user behavior can be used to drive the content changes or to measure the impact of these changes on the user.

\begin{figure*}[t]
  \centering
  \includegraphics[width=1\linewidth]{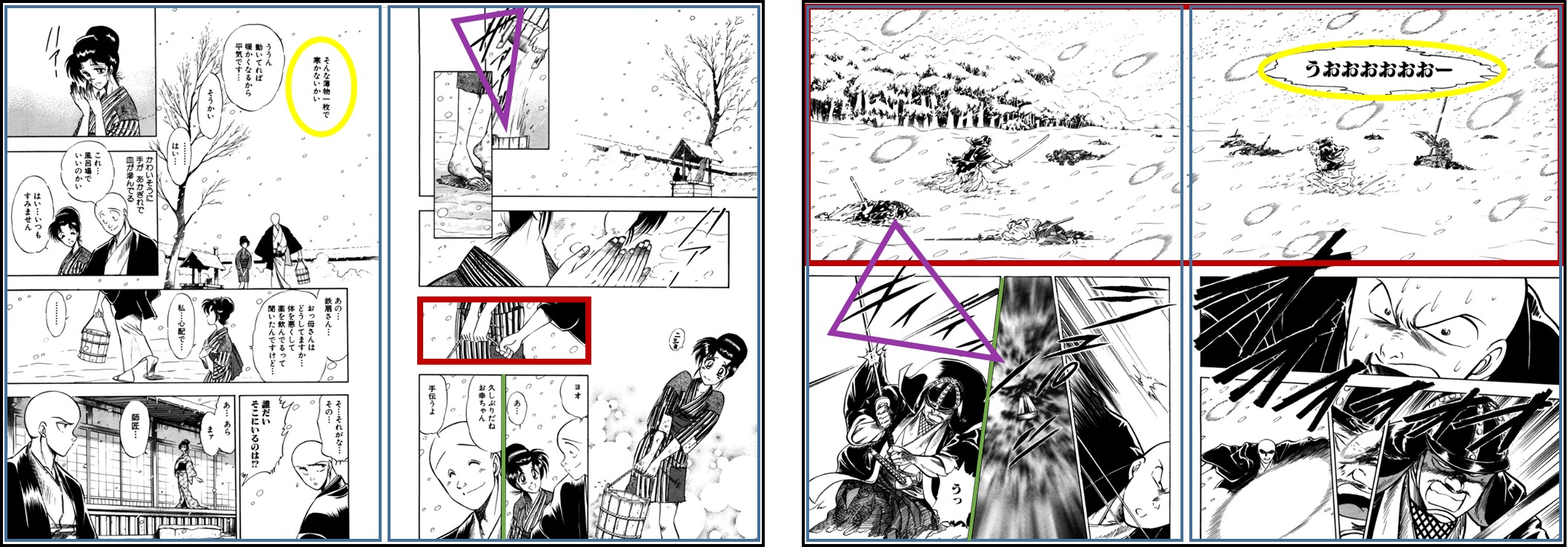}
  \caption{Example of two double comic pages.
  The largest black rectangle encloses two pages which are represented by blue rectangles.
The red rectangles are examples of panel frames. They can be small, large, overlapping each other and have different shapes. Some panels do not have frames and some others can be drawn on more than a page.
The green rectangles are examples of gutters, a white area used for separating two panels.
The yellow ellipses are examples of dialogue balloons. They can have different shapes to represent the feelings of the speaker.
The purple triangles are examples of onomatopoeias, they represent the sound made by people (such as footsteps) or object (water falling, metal blades knocking each other), etc.
Source: images extracted from the Manga109 dataset~\cite{Fujimoto2016}, \textcopyright Sasaki Atsushi. }
  \label{manga}
\end{figure*}

In Section 3 We will state in more detail the current state-of-the-art and discuss the open problems.
Large datasets with ground truth information such as layout, characters, speech balloon, text, etc. are not available so using deep learning is hardly possible in such conditions and most of the researchers proposed handcrafted features or knowledge-driven approaches until the very recent years.
The availability of tools and datasets that can be accessed and shared by the research communities is another very important aspect to transform the research about comics, we will talk about the major existing tools and datasets in Section 4.

In the next parts of the paper all the research which are applied to comics, mangas, bande dessin\'ees or any graphics novels will be referred as ``comics'' in order to simplify the reading.
We start the next section of the paper with general information about comics.

\section{What is comics?}

The term comics (as a singular uncountable noun) refers to the comics medium; such as television, radio, etc. comics is a way to transfer information.
We can also refer to a comic (as a countable noun), in this case, we refer to the instance of the medium such as a comic book or a comic page.  

As for any art, there are strictly no rules for creating comics.
The authors are free to draw whatever and however they want.
Still, some classic layouts or patterns are usually used by the author as they want to tell a story, transmit feelings and emotions, and drive the attention of the readers~\cite{Jain2012}.
The author needs experience and knowledge to drive smoothly the attention of the readers through the comics~\cite{Cao2014}.
Furthermore, the layout of comics is evolving over time~\cite{pederson2016changing}, moving away from conventional grids to a more decorative and dynamic way.

Usually, comics are printed on books and can be seen as a single or double pages.
When the book is opened, the reader can see both pages so some authors use this physical layout as part of the story: some drawings can be spread in two pages, and when the reader turn one page something might happen in the next page.
Figure~\ref{manga} illustrates a classic comics content.

\begin{figure}[t]
  \centering
  \includegraphics[width=0.95\linewidth]{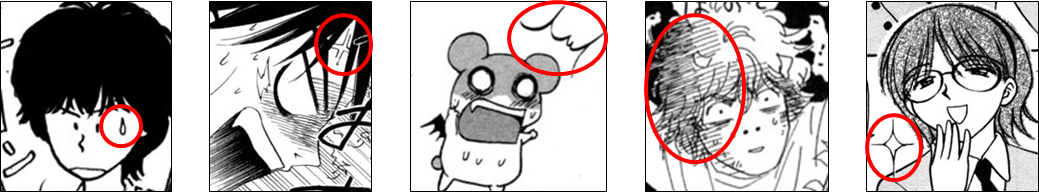}
  \caption{Examples of ``manpu'': a mark used to intensify the emotions of the characters such as concentration, anger, surprise, embarrassment, confidence, etc.
  The original images are extracted from the Manga109 dataset~\cite{Fujimoto2016}, \textcopyright Yoshi Masako, \textcopyright Kobayashi Yuki, \textcopyright Arai Satoshi, \textcopyright Okuda Momoko, \textcopyright Yagami Ken. }
  \label{manpu}
\end{figure}

\begin{figure}[t]
  \centering
  \includegraphics[width=0.3\linewidth]{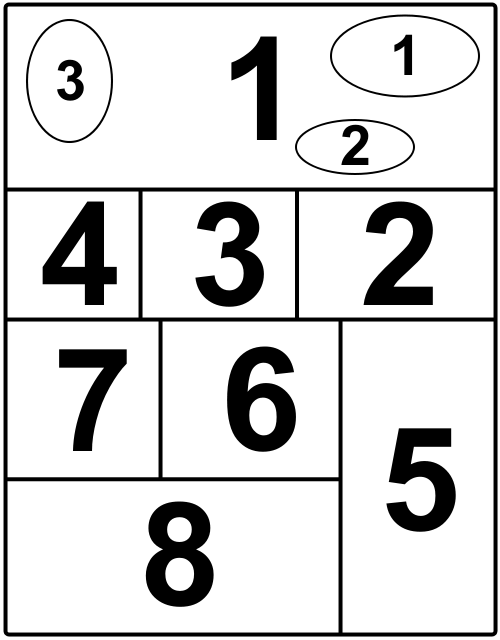}
  \caption{Example of the reading order of a Japanese manga. Image under GNU Free Documentation License: \protect\url{ https://commons.wikimedia.org/wiki/File:Manga_reading_direction.svg}}
  \label{order}
\end{figure}

A page is usually composed of a set of panels defining a specific action or situation.
The panels can be enclosed in a frame and separated by a white space area named gutter.
The reading order of the panels depends on the language.
For example, in Japanese (see Fig.~\ref{order}), the reading order is usually from right to left and top to bottom.
Speech balloons and captions are included in the panel to describe conversations or the narration of the story.
The dialog balloons also have a specified reading order which is usually the same as the reading order of the panels.
Some sound effects or onomatopoeias are often included to give more sensations to the reader such as smell or sound.
Japanese comics often contains ``manpu'' (see Fig.~\ref{manpu}) which are symbols used to visualized feelings and sensations of the characters such as sweating marks on the head of a character to show that he feels uncomfortable even if he is not actually sweating.

\begin{figure}[t]
  \centering
  \includegraphics[width=1\linewidth]{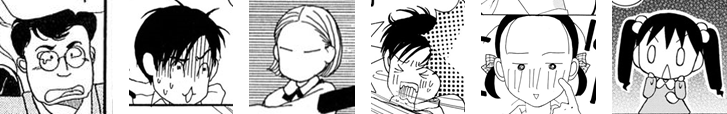}
  \caption{Different examples of comics character faces. Some part of the face such as the nose, eyes, or mouth can be deformed to emphasize the emotion of the character.
The original images are extracted from the Manga109 dataset~\cite{Fujimoto2016}, \textcopyright Kurita Riku, \textcopyright Yamada Uduki, \textcopyright Tenya. }
  \label{faces}
\end{figure}

The authors are free to draw the characters as they want, so they can be deformed or disproportioned as illustrated in Fig.~\ref{faces}. In some genres such as fantasy, the characters can also be non-human which makes the segmentation and recognition task challenging.
There are also many drawing effects such as speed lines, focusing lines, etc.
For example,  in Fig~\ref{manga}, a texture surrounding the female character in the lower-right panel represents her warm atmosphere as contrasted with the cold weather.

\begin{figure}[t]
  \centering
  \includegraphics[width=0.9\linewidth]{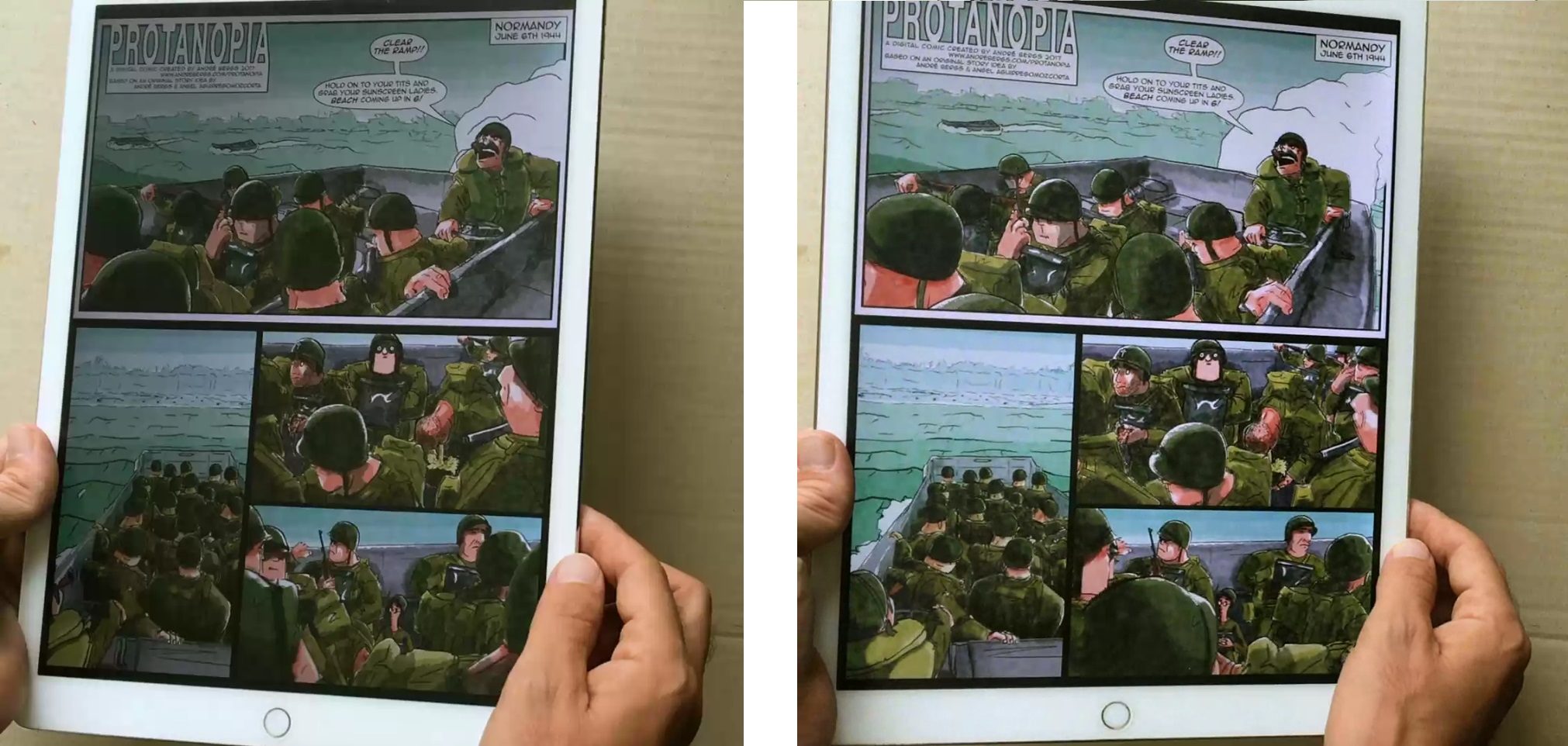}
  \caption{The digital comic ``Protanopia" created by Andre Bergs. The reader can control the camera angle by tilting the screen. The panel are animated with continuous loops. Image extracted from the video on: http://andrebergs.com/protanopia}
  \label{protanopia}
\end{figure}

Even if more and more digitized versions of the printed version are available few comics are produced digitally and taking advantage of the new technology.
Figure~\ref{protanopia} illustrates an example of digital comics taking advantage of tablet functions: the images are animated continuously and the user can tilt the tablet to control the camera angle.
This comics is created by Andre Bergs\footnote{\url{http://andrebergs.com/protanopia}} and is freely available on App store and Google Play.
We imagine that in the future, it could be possible to create such interactive comics automatically.

\section{Comics research}

We organized the studies done about comics in computer science into three main categories that we will present in this section.
One of the main research fields focuses on analyzing the content of comics images, extracting the text, the characters, segmenting the panels, etc.
Another category is about generating new content from or for comics.
The last category is about analyzing the reader's behavior and interaction with comics.

\subsection{Content analysis}

In order to understand the content of comics and to provide services such as retrieval or recommender systems, it is necessary to extract the content of comics.
The DIA community started to cover this problem with classic approaches.
Images can be analyzed from the low levels such as screentones~\cite{ito2015separation} or text~\cite{Arai2011} to the high level such as style~\cite{Chu2016} or genre~\cite{daiku2017comic} recognition.

Some elements are interdependent; for example finding the text and speech balloons, as one can contain the other.
But also the positions can be relative to each other, as the speech balloon is usually coming from the mouth of a character.
These elements are usually grouped inside a panel, but not necessarily.
As the authors are free to draw whatever and however they want, there is a wide disparity among all comics which make the analysis complex.
For example, some authors exaggerate the facial deformation of the face of a character to make him angrier or more surprised.

We present the related work from the low level to high-level analysis as follow.

\subsubsection*{Textures, screentones, and structural lines}
Black and white textures are often used to enrich the visual experience of non-colored comics.
It is especially used for creating an illusion of shades or colors.
However, the identification and segmentation of the textures is challenging as they can have various forms and are sometimes mixed with the other parts of the drawing.
Ito et al. proposed a method for separating the screentones and line drawings~\cite{ito2015separation}.
More recently, Liu et al.~\cite{Liu2017} proposed a method for segmenting the textures in comics.

\begin{figure}[t]
  \centering
  \includegraphics[width=1\linewidth]{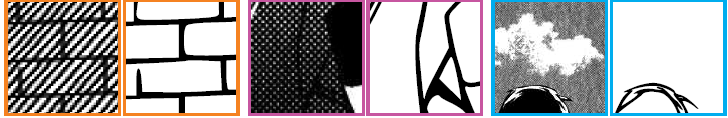}
  \caption{Structural line extraction. For each pair of images, the one on the left is the original image, the one on the right is obtained after removing the textures and detecting the structural lines by Li et al. algorithm~\cite{Li2017}.
Downloaded from: \protect\url{http://exhibition.cintec.cuhk.edu.hk/exhibition/project-item/manga-line-extraction/}. }
  \label{texture}
\end{figure}

Extracting the structural lines of comics is another challenging problem which is related to the analysis of the texture.
The result of such an analysis is displayed in Fig.~\ref{texture}.
The difference between structural lines and arbitrary ones must be considered carefully.
Li et al.~\cite{Li2017} recently proposed a deep network model to handle this problem.
Finding textures and structural lines is an important analysis step to generate colorized and vectorized comics.

\subsubsection*{Text}
The extraction of text (such as Latin or Chinese) characters has been investigated by several researchers but is still a difficult problem as many authors write the text by hand.  

Arai and Tolle~\cite{Arai2011} proposed a method to extract frames, balloon, and text based on connected components and fixed thresholds about their sizes.
This is a simple approach which works well for ``flat'' comics, i.e. conventional comics where each panel is defined by a black rectangle and has no overlapping parts.

Rigaud et al. also proposed a method to recognize the panels and text based on the connected components~\cite{Rigaud2013}.
By adding some other features such as the topological and spatial relations, they successfully increased the performance of~\cite{Arai2011}.

More recently, Aramaki et al. combined connected component and region-based classifications to make a better text detection  system~\cite{Aramaki2016}.
A recent method also addresses the problem of speech text recognition~\cite{rigaud2017segmentation}.

In order to simplify the problem, Hiroe and Hotta have proposed to detect and count the number of exclamation marks in order to represent a comic book by its distribution of exclamation marks or to find the scene changes~\cite{hiroe2017histogram}.

\subsubsection*{Faces and pose}
One of the most important elements of comics is the characters (persons) of the story.
However, identifying the characters is challenging because of the posture, occlusions, and other drawing effects.
Also, the characters can be humans, animals, robots or anything with various drawing representations.
Sun et al.~\cite{Sun2013} proposed to locate and identify the characters in comics pages by using local feature matching.
New methods have recently been proposed to recognize the face and characters in comics based on deep neural networks~\cite{Chu2017,qin2017faster,nguyen2017comic}.

\begin{figure}[t]
  \centering
  \includegraphics[width=1\linewidth]{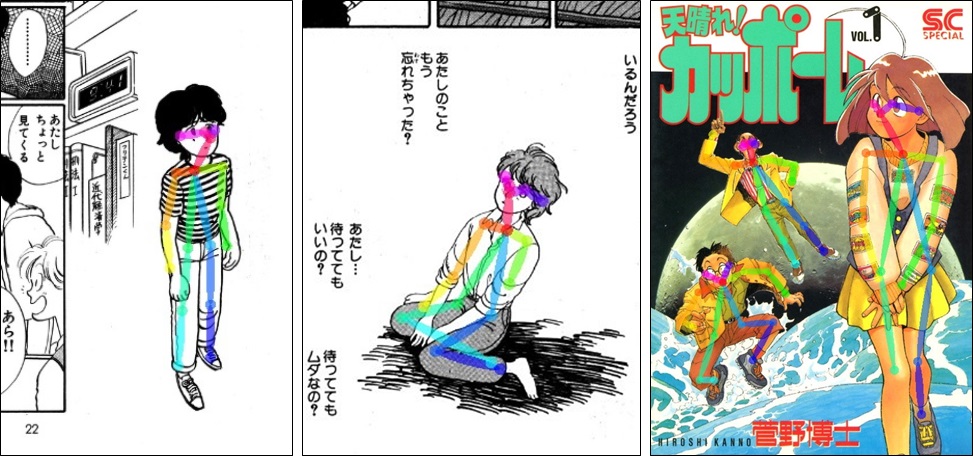}
  \caption{Example of application of Open Pose on comics~\cite{cao2017realtime}. This model works well for comics as long as the drawings are realistic, so it fail in most cases. Source: images extracted from the Manga109 dataset~\cite{Fujimoto2016}, \textcopyright Yoshi Masako, \textcopyright Kanno Hiroshi.}
  \label{pose}
\end{figure}

Estimating the pose of the character is another challenge.
As we can see in Fig.~\ref{pose}, if the characters have human proportion and are not too deformed, they can be well recognized by a popular approach such as Open Pose~\cite{cao2017realtime}.
Knowing the character poses could lead to activity recognition, but a method such as Open Pose will fail on almost all comics.  

\subsubsection*{Balloons}
The balloons are an important component of comics where most of the information is conveyed by the discussion between the protagonists.
So one important step is to detect the balloons~\cite{Correia2016} and then to associate the balloons to the speaker ~\cite{Rigaud2015a}.

The shape of the balloon conveys also information about the speaker feelings~\cite{yamanishi2017speech}. For example, a balloon with wavy shape represents anxiety, an explosion shape represents the anger, a cloudy shape represents joy, etc.

\subsubsection*{Panel}
The layout of a comics page is described by Tanaka et al. as a sequence of frames named panels~\cite{Tanaka2007}.
Several methods have been proposed to segment the panels, mainly based on the analysis of connected components~\cite{Arai2010},~\cite{Rigaud2013} or on the page background mask~\cite{Pang2014}.

As these methods based on heuristics rely on white backgrounds and clean gutters, Iyyer et al. recently proposed an approach based on deep learning~\cite{Iyyer2017} to process eighty-year-old American comics.

\begin{figure}[t]
  \centering
  \includegraphics[width=1\linewidth]{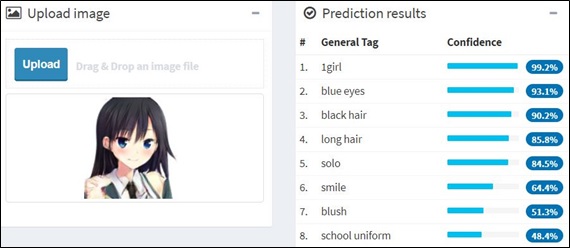}
  \caption{Example application of illustration2vec~\cite{saito2015illustration2vec}. The model recognize several attributes of the character such as her haircut and clothes. The web demo used to generate this image is not online anymore.}
  \label{illust2vec}
\end{figure}

\subsubsection*{High level understanding}
Rigaud et al. proposed a knowledge-driven system that understands the content of comics by segmenting all the sub-parts~\cite{Rigaud2015}.
But understanding the narrative structure of comics is much more than simply segmenting its different sub-parts.
Indeed, the reader makes inferences about what is happening from one frame to another by looking at all graphical and textual elements~\cite{McCloud1993}.

Iyyer et al. introduced some methods to explore how readers connect panels into a coherent story~\cite{Iyyer2017}.
They show that both text and images are important to guess what is happening in a panel by knowing the previous ones.

Daiku et al.~\cite{daiku2017comic} proposed to analyze the comics storytelling by analyzing the genre of each page of the comics.
Then the story of a comic book is represented as a sequence of genres such as: ``11 pages of action'', ``5 pages of romance'', ``8 pages of comedy'', etc.

Analyzing the text of the dialogues and stories has not been investigated yet specifically for comics. Similar research as sentiment analysis~\cite{mohammad2016sentiment} could be applied to analyze the psychology of the characters or to analyze and compare the narrative structure of different comics.

From the cognitive point of view, Cohn proposed a theory of ``Narrative Grammar'' based on linguistics and visual language which are leading the understanding process~\cite{Cohn2013}.
A lot of information is inferred by the reader who is constructing a representation of the depicted pictures in his mind. This is how we can recognize that two characters drawn slightly in a different way are the same, or that a character is doing an action by looking at a still image. These concepts must be inferred by the computer too, in order to obtain a high-level representation of comics.

\subsubsection*{Applications}
From these analyses, retrieval systems can be built, and some have already been proposed in the literature such as sketch~\cite{Matsui2016,narita2017sketch} or graphs based~\cite{Le2016} retrieval.
The drawing style has also been studied~\cite{Chu2016}.
The possible applications are artist retrieval, art movement retrieval, and artwork period analysis.

Saito and Matsui proposed a model for building a feature vector for illustrations named illustration2vec~\cite{saito2015illustration2vec}.
As showed on Fig.\ref{illust2vec}, this model can be used to predict the attributes of a character such as its hair or eye color, the size of the hair, the clothes worn by the character, etc. and to research specific illustrations.
Vie et al. proposed a recommender system using the illustration comics covers based on illustration2vec in a cold-start scenario~\cite{vie2017using}.

\subsubsection*{Conclusion (content analysis)}
Segmenting the panels or reading the text of any comics is still challenging because of the complexity of some layouts and the diversity of the content.
Figure~\ref{manga} illustrates the difficulty of segmenting the panels.
Most of the current methods focus on using handcrafted features for the segmentation and analysis and will fail on an unusual layout.

The segmentation of faces and body of the characters is still an open problem and a large amount of labeled data will be necessary to adapt the deep learning approaches.

Even if the text contains very rich information, surprisingly few methods have been proposed to analyze the storyline or the content of comics based on the text.
Also, some parts of comics has not been addressed at all, such as the detection of onomatopoeias.

The future research about high-level information should be more considered as it can be used to represent information that could interest the reader such as the style or genre, the storytelling, etc.

\subsection{Content generation}

The aim of content generation or enrichment is to use comics to generate new content either based on comics or other media.

\subsubsection*{Vectorization}
As most of comics are not created digitally, vectorization is a way to transform scanned comics to a vector representation for real-time rendering with arbitrary resolution~\cite{Yao2017}.
Generating vectorized comics is necessary for visualizing them nicely in digitized environments.
This is also an important step for editing the content of comics and one of the basic step of comics enrichment~\cite{Zhang2009}.

\subsubsection*{Colorization}
\begin{figure}[t]
  \centering
  \includegraphics[width=0.9\linewidth]{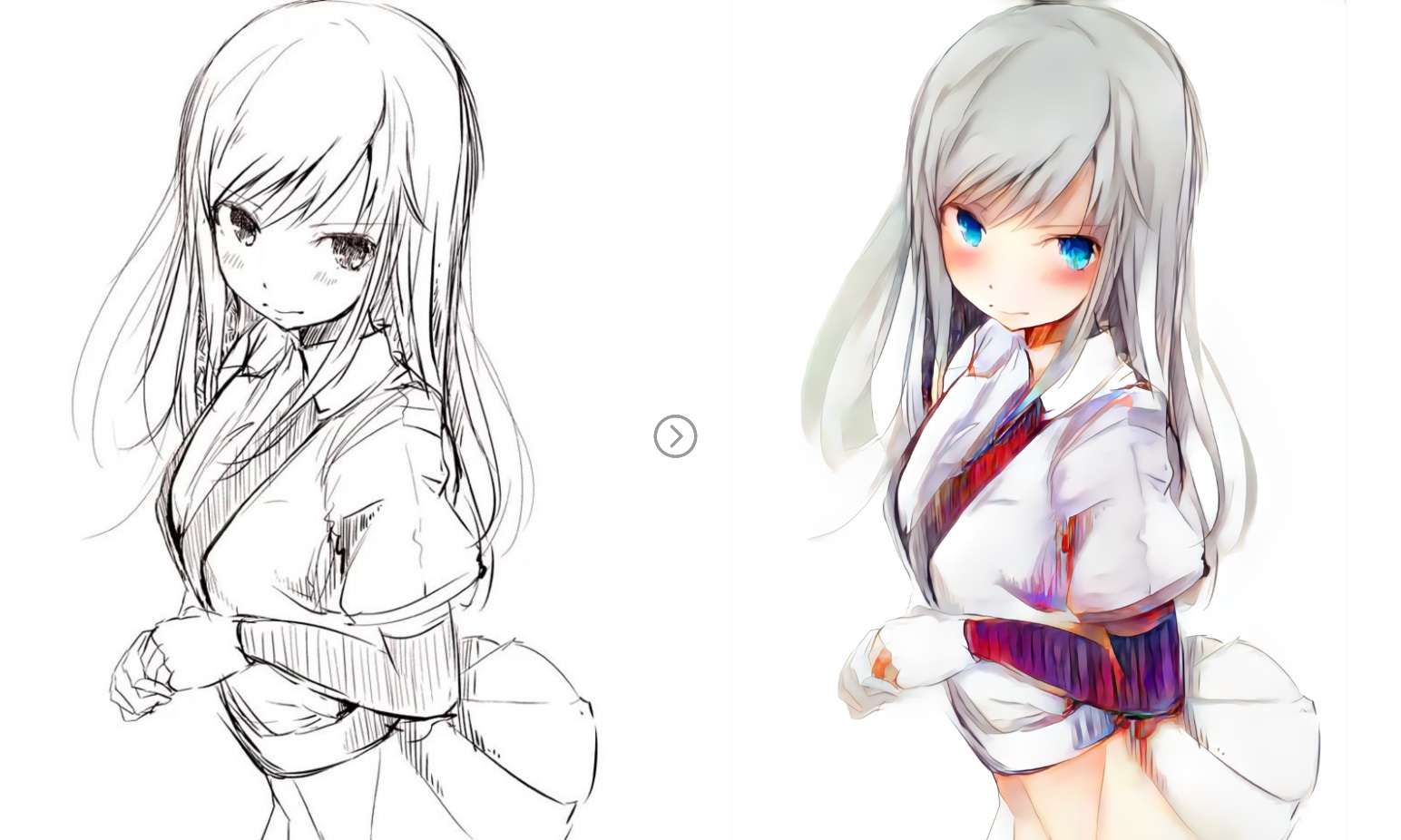}
  \caption{Example of colorization process based on style2paints. Image downloaded from \protect\url{https://github.com/lllyasviel/style2paints}.}
  \label{color}
\end{figure}

Several methods have been proposed for automatic colorization~\cite{Qu2006,Sato2014,cinarelZ17,furusawa2017comicolorization,zhang2017style} and color reconstruction~\cite{Kopf2012}, as comics with colors can be more attractive for some readers.
Colorization is quite a complex problem as the different parts of a character such as his arms, hands, fingers, face, hair, clothes, etc. must be retrieved to color each part in a correct way.
Furthermore, the poses of a character can be very different from each other: some parts can appear, disappear or be deformed.
An example of colorization is displayed in Fig.~\ref{color}.

Recently, deep learning based colorization approach has been used for creating color version manga books which are distributed by professional companies in Japan\footnote{\url{https://www.preferred-networks.jp/en/news/pr20180206}}.

\subsubsection*{Comics and character generation}
One problem for generating comics is to create the layout and to place the different components such as the characters, text balloons, etc. at a correct position to provide a fluid reading experience.  
Cao et al. proposed a method for creating stylistic layout automatically \cite{Cao2012} and then another one for placing and organizing the elements in the panels according to high-level user specification \cite{Cao2014}.

The relation between real-life environment or situations and the one represented in comics can be used to generate or augment comics.
Wu and Aizawa proposed a method to generate a comics image directly from a photograph~\cite{Wu2014}.

\begin{figure}[t]
  \centering
  \includegraphics[width=1\linewidth]{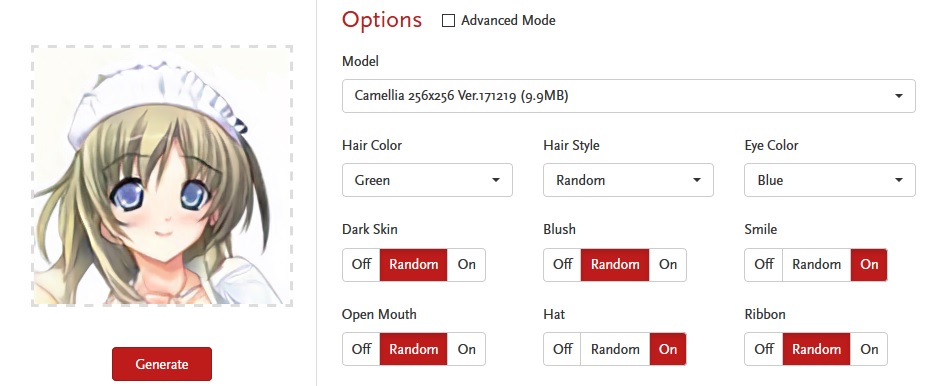}
  \caption{Example of random character generation based Jin et al. method~\cite{jin2017towards}. In this example, we set some attributes such as green hair color, blue eyes, smile and hat.}
  \label{moe}
\end{figure}

At the end of 2017, Jin et al.~\cite{jin2017towards} presented a method to generate automatically comics characters. An example of a generated character by their online demo\footnote{\url{http://make.girls.moe/#/}} is displayed Fig.~\ref{moe}.
The result of the generation is not always visually perfect, but still, this is a powerful tool as an unlimited number of characters can be generated.

\subsubsection*{Animation}
As comics are still images, a way to enhance the visualization of comics is to generate animations.
Recently, some researchers proposed a way for animating still comics images through camera movements~\cite{Cao2017,Jain2016}.
Several animation movies and series have been adapted in comics paper book and vice versa.
Some possible outlook could be to generate an animated movie from a paper comics or a paper comics from an animated movie.

\begin{figure}[t]
  \centering
  \includegraphics[width=1\linewidth]{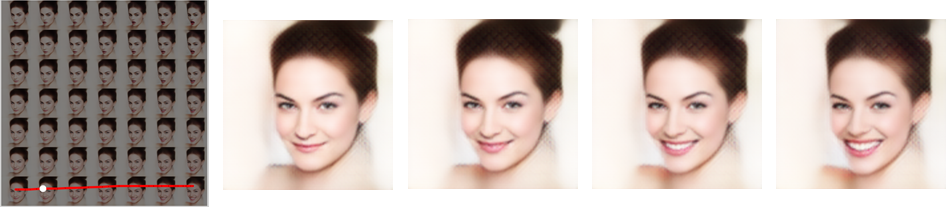}
  \caption{Example of smiling animation in the conceptual space~\cite{white2017generating}. Similar animation could be obtained for comics images. Image source: \protect\url{https://vusd.github.io/toposketch/}}
  \label{smile}
\end{figure}

For the natural images, some methods have been proposed to animate the face of people by using latent space interpolations.
As illustrated in Fig.~\ref{smile} the latent vectors can be computed for a neutral and smiling face to generate a smiling animation~\cite{white2017generating}.

Another application is to use extract the facial keypoints and to use another source (text, speech, or face) to animate the mouth of the character.
For example, this has been done for generating photorealistic video of Obama speech based on a text input~\cite{kumar2017obamanet}.


\subsubsection*{Media conversion}
More broadly, we can imagine to convert text, videos, or any content into comics, and vice-versa.
This problem can be seen as media conversion.
For example, Jing et al. proposed a system to convert videos to comics \cite{Jing2015}.
There are many challenges to do a successful conversion: summarizing the videos, stylizing the images, generating the layout of comics and positions of text balloons.

An application which as not been done to comics but to natural videos is to add generated sound to a video~\cite{zhou2017visual}. No application has been done for comics, but we could imagine a similar application to generate sound effects (swords which are banging to each other, a roaring tailpipe, etc.) or atmosphere sounds (village, countryside, crowd, etc.).

Creating a descriptive text based on comics or generating comics based on descriptive text could be possible in the future, as it has been done for the natural images.
Reed et al.~\cite{reed2016generative} proposed a method for automatic synthesis of realistic natural images from text.

We can also imagine changing the content, adding or removing some parts, changing the genre or style depending on the user or author preference.
 
\subsubsection*{Conclusion (content generation)}

In order to generate contents, some model or labeled data are necessary. In order to generate automatically characters, Jin et al. used around 42000 images.
Deep learning approaches such as Generative Adversarial Networks (GAN)~\cite{goodfellow2014generative} has been widely used for natural image applications such as style transfer~\cite{zhu2017unpaired}, reconstructing 3D models of objects from images~\cite{wu2016learning}, generating images from text~\cite{reed2016generative}, editing pictures~\cite{zhu2016generative}, etc.
These applications could be done for comics too.

Another possibility to enhance comics is to add other modes such as sound, vibrations, etc.
Adding sounds should be easily possible by using the soundtracks from animation movies.  
But, in order to be able to produce these effects at a correct timing, information about the user interactions is necessary.
This is possible by using an eye tracker or detecting when the user turns a specific page in real time.

\subsection{User interaction}

Apart from the content analysis and generation, we have identified another category of research based on the interaction between users and comics.
One part consists of analyzing the user himself instead of analyzing comics.
For example, we would like to understand or predict what the user feels or how he behaves while reading comics.
Another part consists in creating new interfaces or interactions between the readers and comics.
Also, new technology can be used to improve the access for impaired people.

\begin{figure}[t]
  \centering
  \includegraphics[width=0.9\linewidth]{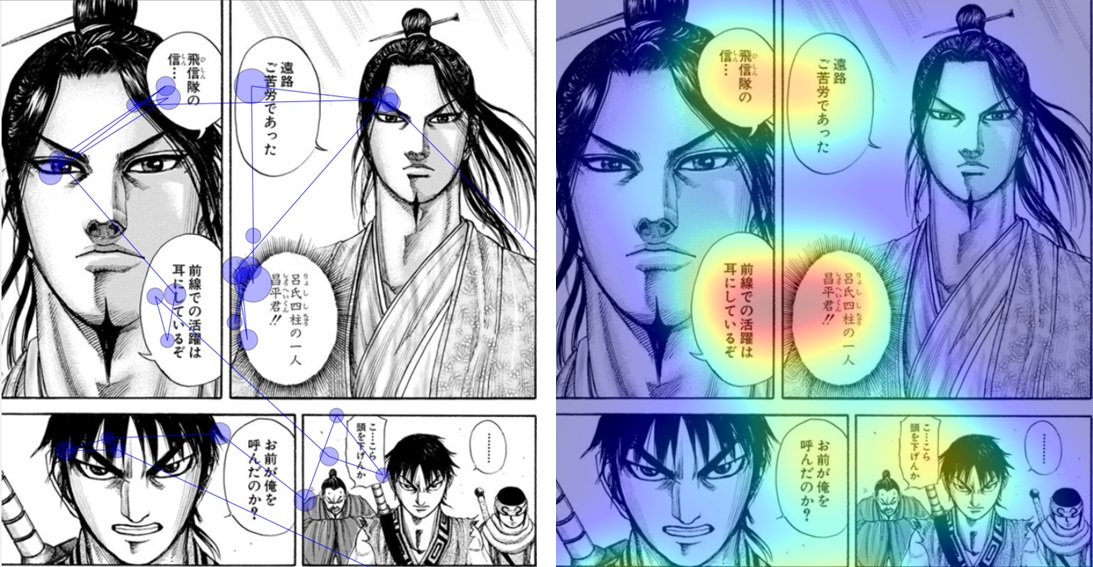}
  \caption{One the left: eye gaze fixations (blue circles) and saccades (segment between circles) of one reader. One the right: heat map accumulated over several readers; the red color corresponds to longer fixation time. }
  \label{eye}
\end{figure}

\subsubsection*{Eye gaze and reading behavior}
In order to know where and when a user is looking at some specific parts of a comic,  researchers are using eye tracking systems.
By using eye trackers it is possible to detect how long a user spends to read a specific part of a comic page.

Knowing the user reading behavior and interest is an important information that can be used by the author or editors as a feedback.
It also can be used to provide other services to readers such as giving more details about the story of a character that a specific user likes, removing part of battle if he does not likes violence, etc.

Carroll et al. ~\cite{carroll1992visual} showed that the readers tend to look at the artworks before reading the text.
Rigaud et al. found that, in France, the readers spend most of the time at reading the text and looking at the face of the characters~\cite{Rigaud2016}. The same experiment repeated in Japan lead to the same conclusion, as illustrated in Fig.~\ref{eye}.

Another way to analyze how the readers understand the content of comics is to ask them to manually order the panels.
Cohn presented different kinds of layouts with empty panels and showed that various manipulations to the arrangement of panels push readers to navigate panels in alternate routes~\cite{cohn2015navigating}.
Some cognitive tricks can ensure that most of the readers will follow the same reading path.

In order to augment comics with new multimedia contents such as sounds, vibration, etc. it is important to trigger these effects at a good timing.
In this case, detecting when the user turns a page or estimating which position he is looking at will be useful.

\subsubsection*{Emotion}
Comics contains exciting contents. Many different genres of comics exist such as comedy, romance, horror, etc. and trigger different kinds of emotions to the readers.
Much research has been done on emotion detection based on face image and physiological signals such as electroencephalogram (EEG) while watching videos~\cite{koelstra2012deap,soleymani2012multimodal,soleymani2016analysis}.
However such research has not been conducted while reading comics.
We think that analyzing the emotion while reading might be more challenging as movie contain animations and sounds that might stimulate more the emotions of the user.

\begin{figure}[t]
  \centering
  \includegraphics[width=0.9\linewidth]{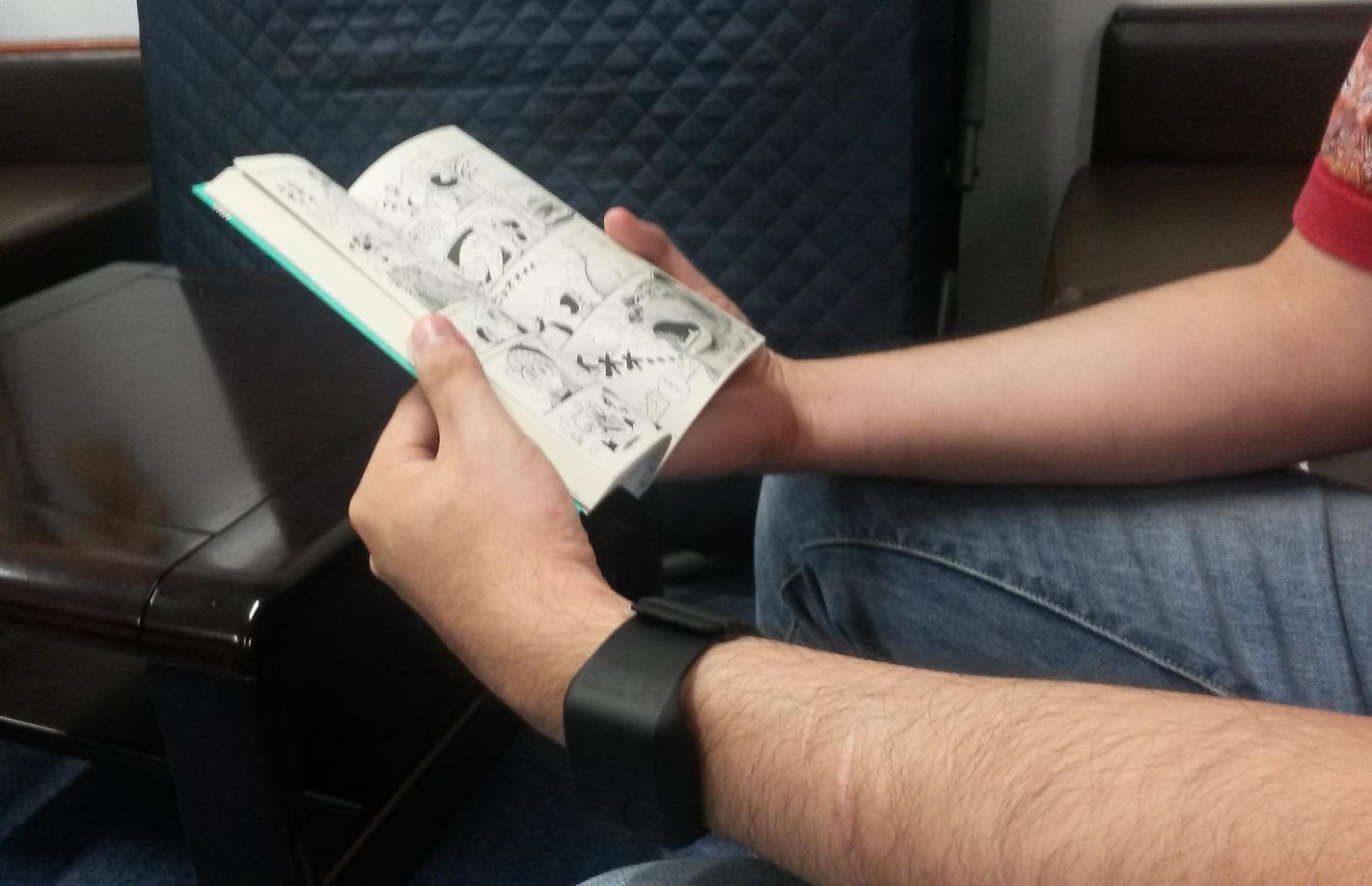}
  \caption{The user wear the E4 wristband, measuring his physiological signals such as heartbeat, skin conductance and skin temperature.}
  \label{e4}
\end{figure}

By recording and analyzing the physiological signals of the readers as illustrated in Fig.~\ref{e4}; Lima Sanches et al. showed that it is possible to estimate if the user is reading a comedy, a romance or a horror comics, based on the emotions felt by the readers~\cite{Sanches2016}. For example, when reading a horror comic book, the user feels stressful and his skin temperature is decreasing.

Emotions are usually represented as two axes: arousal and valence, where the arousal represents the strength of the emotion and the valence relates to a positive or negative emotion.
Matsubara et al. showed that by analyzing the physiological signals of the reader, it is possible to estimate the reader's arousal~\cite{matsubara}.

Both experiments are using the E4 wristband\footnote{\url{https://www.empatica.com/en-eu/research/e4/}} which contains a photoplethysmogram sensor (to analyze the blood volume pulse), an electrodermal activity sensor (to analyze the amount of sweat), an infrared thermopile sensor (to read the peripheral skin temperature), and a 3-axis accelerometer (to captures motion-based activity).
Such device is commonly used for stress detection~\cite{kalimeri2016exploring,greene2016survey}.

Still, each reader has is own preferences and feels emotions in a different way while reading so these analyses are quite challenging.
Depending on the user state of mind or mood, he might prefer to read content that is eliciting specific kind of emotions.
Emotion detection could be used by author or editors to analyze which content stimulate more the readers.

\subsubsection*{Visualization and interaction}
Comics can be read on books, tablets, smartphones or any other devices.  
Visualization and interaction on smartphones can be difficult, especially if the screen is small~\cite{Augereau2016}.
The user needs to zoom and do many operations which can be inconvenient.
Some researchers are also trying to use more interactive devices such as multi-touch tables to attract the users~\cite{Andrews2012}.

Another important challenge is to make comics accessible to impaired people. Rayar~\cite{Rayar} explained that a multidisciplinary collaboration between Human-Computer Interactions, Cognitive Science, and Education Research is necessary to fulfill such a goal.
Up to now, the three main ways to access images for visually impaired people are: audio description, printed Braille description and printed tactile pictures (in relief).
Such way could be generated automatically thanks to new research and technology.

\begin{figure}[t]
  \centering
  \includegraphics[width=0.9\linewidth]{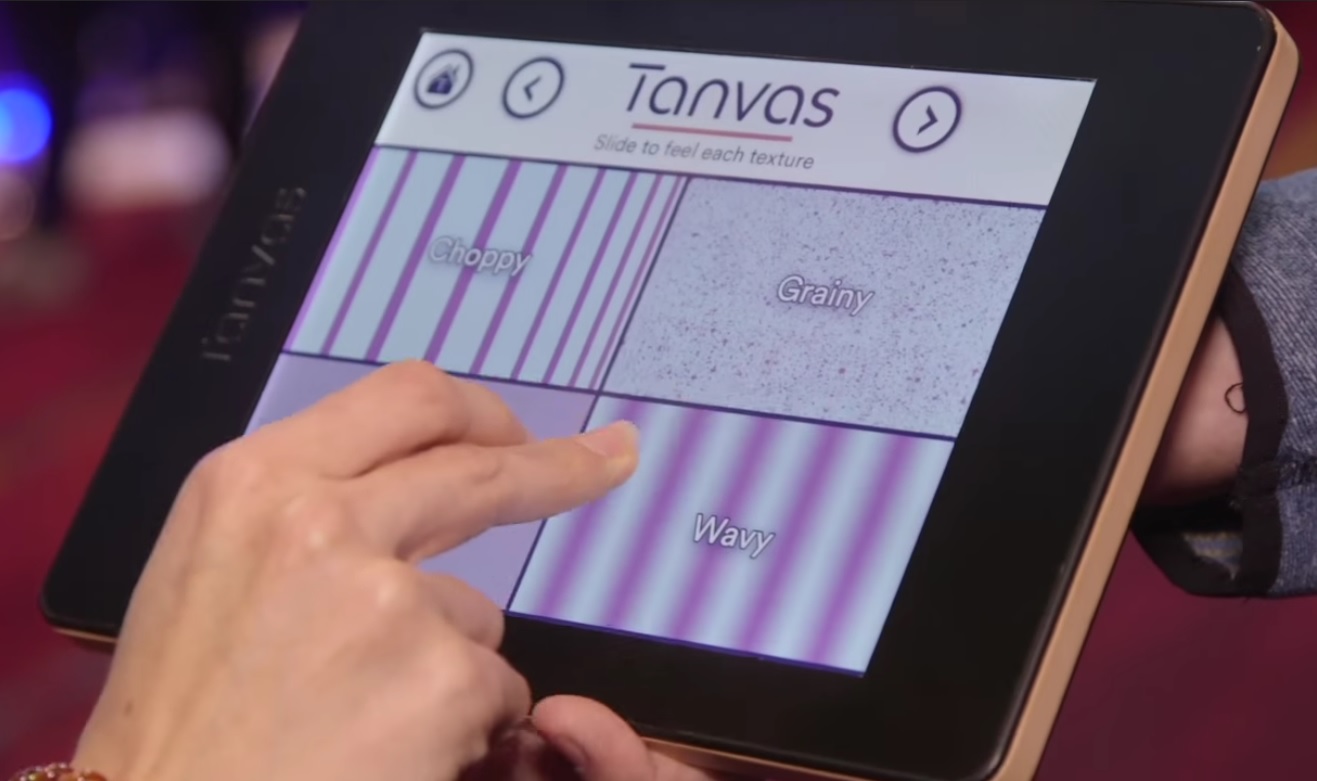}
  \caption{Tanvas tablet enable the user to feel different textures. This could be used to enhance the interaction with comics. Source: \protect\url{https://youtu.be/ohL_B-6Vy6o?t=19s}}
  \label{tanvas}
\end{figure}

New haptic feedback tablet such as the one proposed by Meyer et al.~\cite{meyer2014dynamics} illustrated in Fig.~\ref{tanvas} could help visually impaired people to access comics.
Others application such as detecting and magnifying the text or moving the comics automatically could be helpful for impaired people.

\subsubsection*{Education}
It has been proven that the representation of knowledge as comics can be a good way to attract students to read~\cite{Eneh2008} or to learn language~\cite{Sarada2016}.
It could be interesting to measure the impact on the representation of the knowledge.

Comics could be, for some students, a more interesting way to learn, so using comics in education might be a way to augment their attention level and memory if comics are nicely designed.
A challenge related to media conversion is then to transform normal textbooks into comics and to compare the interactions of the students with both books.

\subsubsection*{Conclusion (user interaction)}
The interactions between the user and comics have not been analyzed deeply yet.
Many sensors can be used to analyze the user with respect to brain activity, muscle activity, body movement and posture, heart rate, sweating, breath, pupil dilation, eye movement, etc.
Collecting such information can give more information about the readers and comics.

\section{Available materials}

In this section, we present some tools and datasets which are publicly available for the research on comics.

\subsection{Tools}
Several tools for comics image segmentation and analysis are available on the Internet and can be freely used by anybody, such as:
 
\begin{itemize}  
\item Speech balloon segmentation~\cite{Rigaud2015b},
\item Speech text recognition~\cite{Rigaud2016a},
\item Automatic text extraction cbrTekStraktor\footnote{\url{https://sourceforge.net/projects/cbrtekstraktor/}},
\item Annotation tool to create ground truth label\footnote{\url{http://www.manga109.org/en/tools/}},
\item Semi-Automatic Manga Colorization~\cite{furusawa2017comicolorization}\footnote{\url{https://github.com/DwangoMediaVillage/Comicolorization}},
\item Deep learning library for estimating a set of tags and extracting semantic feature vectors from illustrations~\cite{saito2015illustration2vec},\footnote{\url{https://github.com/rezoo/illustration2vec}}.
\end{itemize}

The speech balloon~\cite{Rigaud2015b} and text segmentation~\cite{Rigaud2016a} algorithms are available on the author's Github\footnote{\url{https://github.com/crigaud}}.

As we can see, even if many papers have been published about comics segmentation and understanding, still few tools are available on the Internet.
To improve the algorithms significantly and being able to compare them, making the code available is an important step for the community.

\subsection{Datasets}
Few dataset has been made publicly available because of copyright issues.
Indeed, it is not possible for researchers to use and share large dataset of copyrighted materials.
So making competition and reproducible research is not easy.
Hopefully, recently, several datasets have been made available.

The Graphic Narrative Corpus (GNC)~\cite{dunst2017graphic} provide metadata information for 207 titles such as the authors, number of pages, illustrators, genres, etc. Unfortunately, the corresponding images are not available because of copyright protections.
So the usefulness of this dataset is very limited.
Still, the authors are willing to share segmentation ground truth and eye gaze data.
However such data has not been released yet.

eBDtheque~\cite{Guerin2013}\footnote{\url{http://ebdtheque.univ-lr.fr/registration/}} contains 100 comic pages, mainly in French language.
The following elements have been labeled on the dataset: 850 panels, 1092 balloons, 1550 characters and 4691 text lines.
Even if the number of images is limited, creating such detailed labeled data is time-consuming and very useful for the community.

\begin{figure}[t]
  \centering
  \includegraphics[width=1\linewidth]{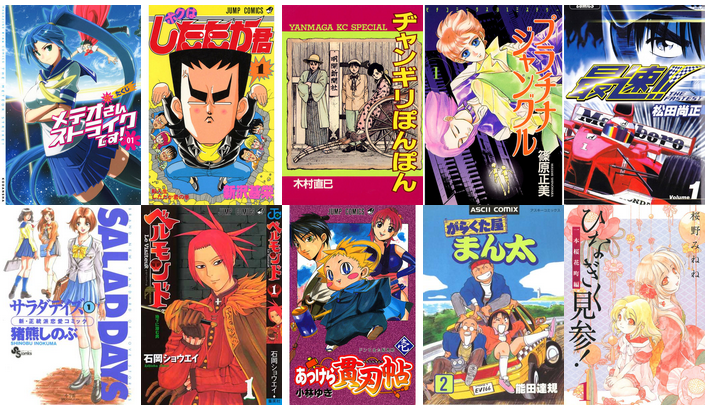}
  \caption{Example of Japanese manga cover page contained in the Manga109 dataset~\cite{Fujimoto2016}.}
  \label{manga109}
\end{figure}

Manga109~\cite{Fujimoto2016}\footnote{\url{http://www.manga109.org/index_en.php}} is illustrated in Fig.~\ref{manga109}.
This dataset which contains 109 manga volumes from 93 different authors.
On average, a volume contains 194 pages.
These mangas were published between the 1970's and 2010's and are categorized into 12 different genres such as fantasy, humor, sports, etc.
Only a limited labeled data are available for now such as the text for few volumes. The strong point of this dataset is to provide all pages of one volume which allows analyzing the sequences of pages.
 
COMICS~\cite{Iyyer2017}\footnote{\url{https://obj.umiacs.umd.edu/comics/index.html}} contains 1,229,664 panels paired with automatic textbox transcriptions from 3,948 American comics books published between 1938 and 1954.
The dataset includes ground truth labeled data such as the rectangular bounding boxes of panels on 500 pages and 1,500 textboxes.

\begin{figure}[t]
  \centering
  \includegraphics[width=1\linewidth]{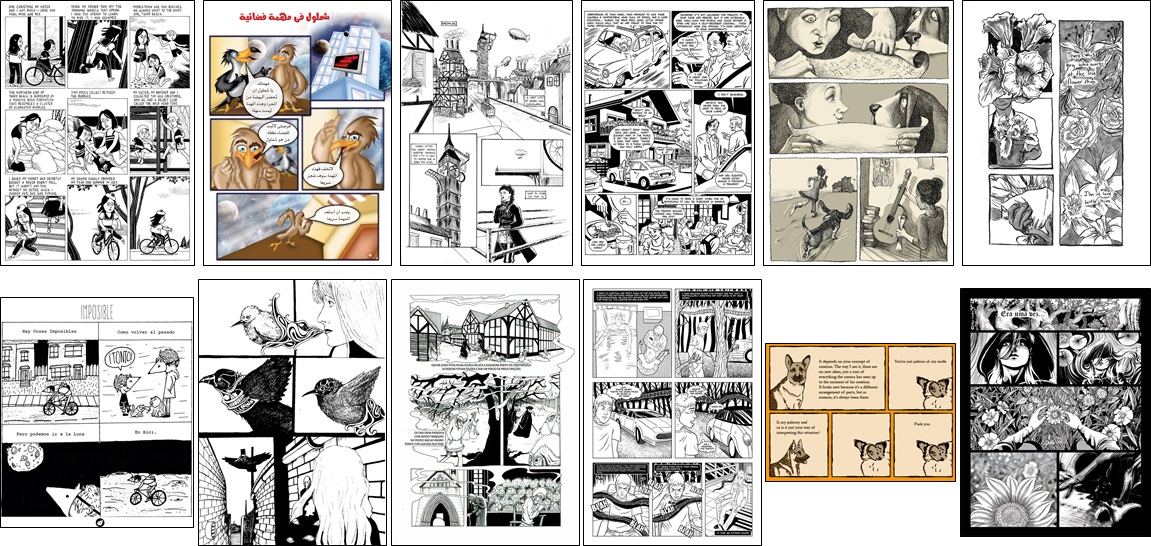}
  \caption{Example of comics contained in the BAM! dataset~\cite{wilber2017bam}. From left to right we selected two images containing the following label: bicycles, birds, buildings, cars, dogs, and flowers. }
  \label{bam}
\end{figure}

BAM!~\cite{wilber2017bam}\footnote{\url{https://bam-dataset.org}}     
contains around 2.5 million artistic images such as: 3D com-
puter graphics, comics, oil painting, pen ink, pencil sketches, vector art, and watercolor. The images contain emotion labels (peaceful, happy,
gloomy, and scary) and object labels (bicycles, birds, buildings, cars, cats,
dogs, flowers, people, and trees).
Figure~\ref{bam} shows a sample of the dataset containing comics.
The dataset is interesting due to the labels and large variety of content and languages. However, the images are just examples provided by the authors and cannot always be understood without the previous or following pages.


BAM!, COMICS, Manga109, and eBDtheques are the four main comics datasets that have been made available with the corresponding images.
Building such datasets is a time and money consuming task, especially for building the ground truth and labeled data.

The main problem to create such dataset comes from the legal and copyright protection which prevent the researchers to make publicly available image datasets.
The content of the dataset is also important depending on the research to proceed. For example, it is interesting to have a variety of comics from different countries, with different languages and genres. It is also interesting to have several continuous pages from the same volumes and several volumes from the same series in order to analyze the evolution of the style of an author, the mentality of the character, or the storyline.

\section{General conclusion}

The research about comics in computer science has been done about several aspects.
We organized the research into three inter-dependent categories: content analysis, content generation, and user interaction.
A mutual analysis of the reader and comics is necessary to understand more about how can we augment comics.

A large part of previous work is focusing on the low-level image analysis by using handcrafted features and knowledge-driven approaches.
Recent research focuses more on deep learning and high-level image understanding.
Still, many applications have been done for natural image and the research about artworks and comics get more attention only very recently~\cite{wilber2017bam}.  

A lot of unexplored fields remain, especially, the content generation and augmentation.
Only few companies started to use research for automatic colorization for example, but it is clear that it could be possible to help the authors with content automatic (or semi-automatic) generation of content or animation.

The analysis of the behavior and emotions of the readers have been superficially covered. However, using the opportunity given by new technologies and sensors could be helpful to create the next age of comics. If could be also a way to help the access of comics to impaired people.

For now, few tools and dataset have been made available. Making publicly available copyrighted images is a problem but it would greatly contribute to the improvement of comics research.

\begin{acknowledgements}
The authors would like to thanks the students of the Intelligent Media Processing Group of Osaka Prefecture University who made some of the presented research and illustrations: Yuki Daiku, Mizuki Matsubara, Charles Lima Sanches, Seiichiro Hara, and Yusuke Maeda.
This work is in part supported by JST CREST (JPMJCR16E1), JSPS Grant-in-Aid for Scientific Research (15K12172), JSPS KAKENHI Grant Number (16K16089) and the Key Project
Grant Program of Osaka Prefecture University.
\end{acknowledgements}

\bibliographystyle{spmpsci}      
\bibliography{manga}

\begin{thebibliography}{10}
\providecommand{\url}[1]{{#1}}
\providecommand{\urlprefix}{URL }
\expandafter\ifx\csname urlstyle\endcsname\relax
  \providecommand{\doi}[1]{DOI~\discretionary{}{}{}#1}\else
  \providecommand{\doi}{DOI~\discretionary{}{}{}\begingroup
  \urlstyle{rm}\Url}\fi

\bibitem{Andrews2012}
Andrews, D., Baber, C., Efremov, S., Komarov, M.: Creating and using
  interactive narratives: reading and writing branching comics.
\newblock In: Proceedings of the SIGCHI Conference on Human Factors in
  Computing Systems, pp. 1703--1712. ACM (2012)

\bibitem{Arai2010}
Arai, K., Tolle, H.: Automatic e-comic content adaptation.
\newblock International Journal of Ubiquitous Computing \textbf{1}(1), 1--11
  (2010)

\bibitem{Arai2011}
Arai, K., Tolle, H.: Method for real time text extraction of digital manga
  comic.
\newblock International Journal of Image Processing (IJIP) \textbf{4}(6),
  669--676 (2011)

\bibitem{Aramaki2016}
Aramaki, Y., Matsui, Y., Yamasaki, T., Aizawa, K.: Text detection in manga by
  combining connected-component-based and region-based classifications.
\newblock In: Image Processing (ICIP), 2016 IEEE International Conference on,
  pp. 2901--2905. IEEE (2016)

\bibitem{augereau2017overview}
Augereau, O., Iwata, M., Kise, K.: An overview of comics research in computer
  science.
\newblock In: 2017 14th IAPR International Conference on Document Analysis and
  Recognition (ICDAR), pp. 54--59. IEEE (2017)

\bibitem{Augereau2016}
Augereau, O., Matsubara, M., Kise, K.: Comic visualization on smartphones based
  on eye tracking.
\newblock In: Proceedings of the 1st International Workshop on coMics ANalysis,
  Processing and Understanding, p.~4. ACM (2016)

\bibitem{Cao2012}
Cao, Y., Chan, A.B., Lau, R.W.: Automatic stylistic manga layout.
\newblock ACM Transactions on Graphics (TOG) \textbf{31}(6), 141 (2012)

\bibitem{Cao2014}
Cao, Y., Lau, R.W., Chan, A.B.: Look over here: Attention-directing composition
  of manga elements.
\newblock ACM Transactions on Graphics (TOG) \textbf{33}(4), 94 (2014)

\bibitem{Cao2017}
Cao, Y., Pang, X., Chan, A.B., Lau, R.W.: Dynamic manga: Animating still manga
  via camera movement.
\newblock IEEE Transactions on Multimedia \textbf{19}(1), 160--172 (2017)

\bibitem{cao2017realtime}
Cao, Z., Simon, T., Wei, S.E., Sheikh, Y.: Realtime multi-person 2d pose
  estimation using part affinity fields.
\newblock In: CVPR, vol.~1, p.~7 (2017)

\bibitem{carroll1992visual}
Carroll, P.J., Young, J.R., Guertin, M.S.: Visual analysis of cartoons: A view
  from the far side.
\newblock In: Eye movements and visual cognition, pp. 444--461. Springer (1992)

\bibitem{Christiansen2000}
Christiansen, H.C.: Comics \& culture: analytical and theoretical approaches to
  comics.
\newblock Museum Tusculanum Press (2000)

\bibitem{Chu2016}
Chu, W.T., Cheng, W.C.: Manga-specific features and latent style model for
  manga style analysis.
\newblock In: Acoustics, Speech and Signal Processing (ICASSP), 2016 IEEE
  International Conference on, pp. 1332--1336. IEEE (2016)

\bibitem{Chu2017}
Chu, W.T., Li, W.W.: Manga facenet: Face detection in manga based on deep
  neural network.
\newblock In: Proceedings of the 2017 ACM on International Conference on
  Multimedia Retrieval, pp. 412--415. ACM (2017)

\bibitem{cinarelZ17}
Cinarel, C., Zhang, B.: Into the colorful world of webtoons: Through the lens
  of neural networks.
\newblock In: 2nd International Workshop on coMics Analysis, Processing, and
  Understanding, 14th {IAPR} International Conference on Document Analysis and
  Recognition, {ICDAR} 2017, Kyoto, Japan, November 9-15, 2017, pp. 35--40
  (2017)

\bibitem{Cohn2013}
Cohn, N.: Visual narrative structure.
\newblock Cognitive science \textbf{37}(3), 413--452 (2013)

\bibitem{cohn2015navigating}
Cohn, N., Campbell, H.: Navigating comics ii: Constraints on the reading order
  of comic page layouts.
\newblock Applied Cognitive Psychology \textbf{29}(2), 193--199 (2015)

\bibitem{Correia2016}
Correia, J.M., Gomes, A.J.: Balloon extraction from complex comic books using
  edge detection and histogram scoring.
\newblock Multimedia Tools and Applications \textbf{75}(18), 11367--11390
  (2016)

\bibitem{daiku2017comic}
Daiku, Y., Augereau, O., Iwata, M., Kise, K.: Comic story analysis based on
  genre classification.
\newblock In: 2017 14th IAPR International Conference on Document Analysis and
  Recognition (ICDAR), pp. 60--65. IEEE (2017)

\bibitem{dunst2017graphic}
Dunst, A., Hartel, R., Laubrock, J.: The graphic narrative corpus (gnc):
  Design, annotation, and analysis for the digital humanities.
\newblock In: 2017 14th IAPR International Conference on Document Analysis and
  Recognition (ICDAR), pp. 15--20. IEEE (2017)

\bibitem{Eneh2008}
Eneh, A., Eneh, O.: Enhancing pupils' reading achievement by use of comics and
  cartoons in teaching reading.
\newblock Journal of Applied Science \textbf{11}(3), 8058--62 (2008)

\bibitem{Fujimoto2016}
Fujimoto, A., Ogawa, T., Yamamoto, K., Matsui, Y., Yamasaki, T., Aizawa, K.:
  Manga109 dataset and creation of metadata.
\newblock In: Proceedings of the 1st International Workshop on coMics ANalysis,
  Processing and Understanding, p.~2. ACM (2016)

\bibitem{furusawa2017comicolorization}
Furusawa, C., Hiroshiba, K., Ogaki, K., Odagiri, Y.: Comicolorization:
  semi-automatic manga colorization.
\newblock In: SIGGRAPH Asia 2017 Technical Briefs, p.~12. ACM (2017)

\bibitem{goodfellow2014generative}
Goodfellow, I., Pouget-Abadie, J., Mirza, M., Xu, B., Warde-Farley, D., Ozair,
  S., Courville, A., Bengio, Y.: Generative adversarial nets.
\newblock In: Advances in neural information processing systems, pp. 2672--2680
  (2014)

\bibitem{greene2016survey}
Greene, S., Thapliyal, H., Caban-Holt, A.: A survey of affective computing for
  stress detection: Evaluating technologies in stress detection for better
  health.
\newblock IEEE Consumer Electronics Magazine \textbf{5}(4), 44--56 (2016)

\bibitem{Guerin2013}
Gu{\'e}rin, C., Rigaud, C., Mercier, A., Ammar-Boudjelal, F., Bertet, K.,
  Bouju, A., Burie, J.C., Louis, G., Ogier, J.M., Revel, A.: ebdtheque: a
  representative database of comics.
\newblock In: Document Analysis and Recognition (ICDAR), 2013 12th
  international conference on, pp. 1145--1149. IEEE (2013)

\bibitem{hall2013struggle}
Hall, I., Smith, F.: The struggle for soft power in asia: Public diplomacy and
  regional competition.
\newblock Asian Security \textbf{9}(1), 1--18 (2013)

\bibitem{hiroe2017histogram}
Hiroe, S., Hotta, S.: Histogram of exclamation marks and its application for
  comics analysis.
\newblock In: 2017 14th IAPR International Conference on Document Analysis and
  Recognition (ICDAR), pp. 66--71. IEEE (2017)

\bibitem{ito2015separation}
Ito, K., Matsui, Y., Yamasaki, T., Aizawa, K.: Separation of manga line
  drawings and screentones.
\newblock In: Eurographics (Short Papers), pp. 73--76 (2015)

\bibitem{Iyyer2017}
Iyyer, M., Manjunatha, V., Guha, A., Vyas, Y., Boyd-Graber, J., {Daum\'{e}
  III}, H., Davis, L.: The amazing mysteries of the gutter: Drawing inferences
  between panels in comic book narratives.
\newblock In: IEEE Conference on Computer Vision and Pattern Recognition (2017)

\bibitem{Jain2012}
Jain, E., Sheikh, Y., Hodgins, J.: Inferring artistic intention in comic art
  through viewer gaze.
\newblock In: Proceedings of the ACM Symposium on Applied Perception, pp.
  55--62. ACM (2012)

\bibitem{Jain2016}
Jain, E., Sheikh, Y., Hodgins, J.: Predicting moves-on-stills for comic art
  using viewer gaze data.
\newblock IEEE Computer Graphics and Applications \textbf{36}(4), 34--45 (2016)

\bibitem{jin2017towards}
Jin, Y., Zhang, J., Li, M., Tian, Y., Zhu, H., Fang, Z.: Towards the automatic
  anime characters creation with generative adversarial networks.
\newblock arXiv preprint arXiv:1708.05509  (2017)

\bibitem{Jing2015}
Jing, G., Hu, Y., Guo, Y., Yu, Y., Wang, W.: Content-aware video2comics with
  manga-style layout.
\newblock IEEE Transactions on Multimedia \textbf{17}(12), 2122--2133 (2015)

\bibitem{kalimeri2016exploring}
Kalimeri, K., Saitis, C.: Exploring multimodal biosignal features for stress
  detection during indoor mobility.
\newblock In: Proceedings of the 18th ACM International Conference on
  Multimodal Interaction, pp. 53--60. ACM (2016)

\bibitem{koelstra2012deap}
Koelstra, S., Muhl, C., Soleymani, M., Lee, J.S., Yazdani, A., Ebrahimi, T.,
  Pun, T., Nijholt, A., Patras, I.: Deap: A database for emotion analysis;
  using physiological signals.
\newblock IEEE Transactions on Affective Computing \textbf{3}(1), 18--31 (2012)

\bibitem{Kopf2012}
Kopf, J., Lischinski, D.: Digital reconstruction of halftoned color comics.
\newblock ACM Transactions on Graphics (TOG) \textbf{31}(6), 140 (2012)

\bibitem{kumar2017obamanet}
Kumar, R., Sotelo, J., Kumar, K., de~Brebisson, A., Bengio, Y.: Obamanet:
  Photo-realistic lip-sync from text.
\newblock arXiv preprint arXiv:1801.01442  (2017)

\bibitem{lam2007japan}
Lam, P.E.: Japan’s quest for “soft power”: attraction and limitation.
\newblock East Asia \textbf{24}(4), 349--363 (2007)

\bibitem{Le2016}
Le, T.N., Luqman, M.M., Burie, J.C., Ogier, J.M.: Retrieval of comic book
  images using context relevance information.
\newblock In: Proceedings of the 1st International Workshop on coMics ANalysis,
  Processing and Understanding, p.~12. ACM (2016)

\bibitem{Li2017}
Li, C., Liu, X., Wong, T.T.: Deep extraction of manga structural lines.
\newblock ACM Transactions on Graphics (TOG) \textbf{36}(4), 117 (2017)

\bibitem{Sanches2016}
Lima~Sanches, C., Augereau, O., Kise, K.: Manga content analysis using
  physiological signals.
\newblock In: Proceedings of the 1st International Workshop on coMics ANalysis,
  Processing and Understanding, p.~6. ACM (2016)

\bibitem{Liu2017}
Liu, X., Li, C., Wong, T.T.: Boundary-aware texture region segmentation from
  manga.
\newblock Computational Visual Media \textbf{3}(1), 61--71 (2017)

\bibitem{matsubara}
Matsubara, M., Augereau, O., Sanches, C.L., Kise, K.: Emotional arousal
  estimation while reading comics based on physiological signal analysis.
\newblock In: Proceedings of the 1st International Workshop on coMics ANalysis,
  Processing and Understanding, MANPU '16, pp. 7:1--7:4. ACM, New York, NY, USA
  (2016)

\bibitem{Matsui2016}
Matsui, Y., Ito, K., Aramaki, Y., Fujimoto, A., Ogawa, T., Yamasaki, T.,
  Aizawa, K.: Sketch-based manga retrieval using manga109 dataset.
\newblock Multimedia Tools and Applications pp. 1--28 (2016)

\bibitem{McCloud1993}
McCloud, S.: Understanding comics: The invisible art.
\newblock Northampton, Mass  (1993)

\bibitem{meyer2014dynamics}
Meyer, D.J., Wiertlewski, M., Peshkin, M.A., Colgate, J.E.: Dynamics of
  ultrasonic and electrostatic friction modulation for rendering texture on
  haptic surfaces.
\newblock In: Haptics Symposium (HAPTICS), 2014 IEEE, pp. 63--67. IEEE (2014)

\bibitem{mohammad2016sentiment}
Mohammad, S.M.: Sentiment analysis: Detecting valence, emotions, and other
  affectual states from text.
\newblock In: Emotion measurement, pp. 201--237. Elsevier (2016)

\bibitem{narita2017sketch}
Narita, R., Tsubota, K., Yamasaki, T., Aizawa, K.: Sketch-based manga retrieval
  using deep features.
\newblock In: 2017 14th IAPR International Conference on Document Analysis and
  Recognition (ICDAR), pp. 49--53. IEEE (2017)

\bibitem{nguyen2017comic}
Nguyen, N.V., Rigaud, C., Burie, J.C.: Comic characters detection using deep
  learning.
\newblock In: 2017 14th IAPR International Conference on Document Analysis and
  Recognition (ICDAR), pp. 41--46. IEEE (2017)

\bibitem{Pang2014}
Pang, X., Cao, Y., Lau, R.W., Chan, A.B.: A robust panel extraction method for
  manga.
\newblock In: Proceedings of the 22nd ACM international conference on
  Multimedia, pp. 1125--1128. ACM (2014)

\bibitem{pederson2016changing}
Pederson, K., Cohn, N.: The changing pages of comics: Page layouts across eight
  decades of american superhero comics.
\newblock Studies in Comics \textbf{7}(1), 7--28 (2016)

\bibitem{qin2017faster}
Qin, X., Zhou, Y., He, Z., Wang, Y., Tang, Z.: A faster r-cnn based method for
  comic characters face detection.
\newblock In: 2017 14th IAPR International Conference on Document Analysis and
  Recognition (ICDAR), pp. 1074--1080. IEEE (2017)

\bibitem{Qu2006}
Qu, Y., Wong, T.T., Heng, P.A.: Manga colorization.
\newblock In: ACM Transactions on Graphics (TOG), vol.~25, pp. 1214--1220. ACM
  (2006)

\bibitem{Rayar}
Rayar, F.: Accessible comics for visually impaired people: Challenges and
  opportunities.
\newblock In: 2017 14th IAPR International Conference on Document Analysis and
  Recognition (ICDAR), vol.~03, pp. 9--14 (2017).
\newblock \doi{10.1109/ICDAR.2017.285}

\bibitem{reed2016generative}
Reed, S., Akata, Z., Yan, X., Logeswaran, L., Schiele, B., Lee, H.: Generative
  adversarial text to image synthesis.
\newblock arXiv preprint arXiv:1605.05396  (2016)

\bibitem{Rigaud2015b}
Rigaud, C., Burie, J.C., Ogier, J.M.: Text-Independent Speech Balloon
  Segmentation for Comics and Manga, pp. 133--147.
\newblock Springer International Publishing, Cham (2015)

\bibitem{rigaud2017segmentation}
Rigaud, C., Burie, J.C., Ogier, J.M.: Segmentation-free speech text recognition
  for comic books.
\newblock In: 2017 14th IAPR International Conference on Document Analysis and
  Recognition (ICDAR), pp. 29--34. IEEE (2017)

\bibitem{Rigaud2015}
Rigaud, C., Gu{\'e}rin, C., Karatzas, D., Burie, J.C., Ogier, J.M.:
  Knowledge-driven understanding of images in comic books.
\newblock International Journal on Document Analysis and Recognition (IJDAR)
  \textbf{18}(3), 199--221 (2015)

\bibitem{Rigaud2016}
Rigaud, C., Le, T.N., Burie, J.C., Ogier, J.M., Ishimaru, S., Iwata, M., Kise,
  K.: Semi-automatic text and graphics extraction of manga using eye tracking
  information.
\newblock In: Document Analysis Systems (DAS), 2016 12th IAPR Workshop on, pp.
  120--125. IEEE (2016)

\bibitem{Rigaud2015a}
Rigaud, C., Le~Thanh, N., Burie, J.C., Ogier, J.M., Iwata, M., Imazu, E., Kise,
  K.: Speech balloon and speaker association for comics and manga
  understanding.
\newblock In: Document Analysis and Recognition (ICDAR), 2015 13th
  International Conference on, pp. 351--355. IEEE (2015)

\bibitem{Rigaud2016a}
Rigaud, C., Pal, S., Burie, J.C., Ogier, J.M.: Toward speech text recognition
  for comic books.
\newblock In: Proceedings of the 1st International Workshop on coMics ANalysis,
  Processing and Understanding, MANPU '16, pp. 8:1--8:6. ACM, New York, NY, USA
  (2016).
\newblock \doi{10.1145/3011549.3011557}.
\newblock \urlprefix\url{http://doi.acm.org/10.1145/3011549.3011557}

\bibitem{Rigaud2013}
Rigaud, C., Tsopze, N., Burie, J.C., Ogier, J.M.: Robust frame and text
  extraction from comic books.
\newblock In: Graphics Recognition. New Trends and Challenges, pp. 129--138.
  Springer (2013)

\bibitem{saito2015illustration2vec}
Saito, M., Matsui, Y.: Illustration2vec: a semantic vector representation of
  illustrations.
\newblock In: SIGGRAPH Asia 2015 Technical Briefs, p.~5. ACM (2015)

\bibitem{Sarada2016}
Sarada, P.: Comics as a powerful tool to enhance english language usage.
\newblock IUP Journal of English Studies \textbf{11}(1), 60 (2016)

\bibitem{Sato2014}
Sato, K., Matsui, Y., Yamasaki, T., Aizawa, K.: Reference-based manga
  colorization by graph correspondence using quadratic programming.
\newblock In: SIGGRAPH Asia 2014 Technical Briefs, p.~15. ACM (2014)

\bibitem{Screech2005}
Screech, M.: Masters of the ninth art: bandes dessin{\'e}es and Franco-Belgian
  identity, vol.~3.
\newblock Liverpool University Press (2005)

\bibitem{soleymani2016analysis}
Soleymani, M., Asghari-Esfeden, S., Fu, Y., Pantic, M.: Analysis of eeg signals
  and facial expressions for continuous emotion detection.
\newblock IEEE Transactions on Affective Computing \textbf{7}(1), 17--28 (2016)

\bibitem{soleymani2012multimodal}
Soleymani, M., Lichtenauer, J., Pun, T., Pantic, M.: A multimodal database for
  affect recognition and implicit tagging.
\newblock IEEE Transactions on Affective Computing \textbf{3}(1), 42--55 (2012)

\bibitem{Sun2013}
Sun, W., Burie, J.C., Ogier, J.M., Kise, K.: Specific comic character detection
  using local feature matching.
\newblock In: Document Analysis and Recognition (ICDAR), 2013 12th
  International Conference on, pp. 275--279. IEEE (2013)

\bibitem{Tanaka2007}
Tanaka, T., Shoji, K., Toyama, F., Miyamichi, J.: Layout analysis of
  tree-structured scene frames in comic images.
\newblock In: IJCAI, vol.~7, pp. 2885--2890 (2007)

\bibitem{vie2017using}
Vie, J.J., Yger, F., Lahfa, R., Clement, B., Cocchi, K., Chalumeau, T.,
  Kashima, H.: Using posters to recommend anime and mangas in a cold-start
  scenario.
\newblock arXiv preprint arXiv:1709.01584  (2017)

\bibitem{white2017generating}
White, T., Loh, I.: Generating animations by sketching in conceptual space.
\newblock In: Eighth International Conference on Computational Creativity,
  ICCC, Atlanta (2017)

\bibitem{wilber2017bam}
Wilber, M.J., Fang, C., Jin, H., Hertzmann, A., Collomosse, J., Belongie, S.:
  Bam! the behance artistic media dataset for recognition beyond photography.
\newblock In: Proc. ICCV, vol.~1, p.~4 (2017)

\bibitem{wu2016learning}
Wu, J., Zhang, C., Xue, T., Freeman, B., Tenenbaum, J.: Learning a
  probabilistic latent space of object shapes via 3d generative-adversarial
  modeling.
\newblock In: Advances in Neural Information Processing Systems, pp. 82--90
  (2016)

\bibitem{Wu2014}
Wu, Z., Aizawa, K.: Mangawall: Generating manga pages for real-time
  applications.
\newblock In: Acoustics, Speech and Signal Processing (ICASSP), 2014 IEEE
  International Conference on, pp. 679--683. IEEE (2014)

\bibitem{yamanishi2017speech}
Yamanishi, R., Tanaka, H., Nishihara, Y., Fukumoto, J.: Speech-balloon shapes
  estimation for emotional text communication.
\newblock Information Engineering Express \textbf{3}(2), 1--10 (2017)

\bibitem{Yao2017}
Yao, C.Y., Hung, S.H., Li, G.W., Chen, I.Y., Adhitya, R., Lai, Y.C.: Manga
  vectorization and manipulation with procedural simple screentone.
\newblock IEEE transactions on visualization and computer graphics
  \textbf{23}(2), 1070--1084 (2017)

\bibitem{zhang2017style}
Zhang, L., Ji, Y., Lin, X.: Style transfer for anime sketches with enhanced
  residual u-net and auxiliary classifier gan.
\newblock arXiv preprint arXiv:1706.03319  (2017)

\bibitem{Zhang2009}
Zhang, S.H., Chen, T., Zhang, Y.F., Hu, S.M., Martin, R.R.: Vectorizing cartoon
  animations.
\newblock IEEE Transactions on Visualization and Computer Graphics
  \textbf{15}(4), 618--629 (2009)

\bibitem{zhou2017visual}
Zhou, Y., Wang, Z., Fang, C., Bui, T., Berg, T.L.: Visual to sound: Generating
  natural sound for videos in the wild.
\newblock arXiv preprint arXiv:1712.01393  (2017)

\bibitem{zhu2016generative}
Zhu, J.Y., Kr{\"a}henb{\"u}hl, P., Shechtman, E., Efros, A.A.: Generative
  visual manipulation on the natural image manifold.
\newblock In: European Conference on Computer Vision, pp. 597--613. Springer
  (2016)

\bibitem{zhu2017unpaired}
Zhu, J.Y., Park, T., Isola, P., Efros, A.A.: Unpaired image-to-image
  translation using cycle-consistent adversarial networks.
\newblock arXiv preprint arXiv:1703.10593  (2017)

\end{thebibliography}

\end{document}